\documentstyle[twoside,jltp]{article}

\title{The Pairing Mechanism in HTSC investigated by Electronic Raman
Scattering}

\author{A. Hoffmann, P. Lemmens, L. Winkeler, and G. G\"untherodt
\address{2. Physikalisches Institut der RWTH-Aachen, 52056 Aachen, Germany}}

\runninghead{A. Hoffmann, P. Lemmens, L. Winkeler, and G. G\"untherodt}
{The Pairing Mechanism in HTSC investigated by Electronic Raman Scattering}

\begin{document}

\begin{abstract}
By means of electronic Raman scattering we investigated the symmetry of the
energy gap $2\Delta(\vec{k})$, its temperature dependence and its variation
with doping of well characterized Bi$_2$Sr$_2$CaCu$_2$O$_{8+\delta}$ single
crystals. The oxygen content $\delta$ was varied between the under- and the
overdoped regime by subsequently annealing the same single crystal in Ar and
in O$_2$, respectively. The symmetry analysis of the polarized electronic
Raman scattering is consistent with a d$_{x^2-y^2}$-wave symmetry of the
energy gap in both regimes. The gap ratio $2\Delta_{max}/k_BT_c$ and its
temperature dependence changes with doping similar to predictions of theories
based on paramagnon coupling.
\end{abstract}

\maketitle

\section{INTRODUCTION}
The symmetry of the superconducting order parameter or energy gap
$2\Delta(\vec{k})$ gives an important clue to the mechanism of high T$_c$
superconductivity. Experiments which probe a possible phase shift in the
order parameter with Josphson tunneling\cite{Schrieffer} are in favor of a
d$_{x^2-y^2}$-wave pairing. Moreover, various experiments\cite{Schrieffer},
including electronic Raman scattering\cite{Devereaux}, are indicating the
existence of nodes in the energy gap. A d$_{x^2-y^2}$-symmetry of the order
parameter is predicted for a pairing mechanism based on spin
fluctuations\cite{Schrieffer}. Furthermore, there are theoretical predictions
for the magnitude of the energy gap and its temperature dependence\cite{Pao}.

Electronic Raman scattering of free carriers occurs due to mass fluctuations
about the Fermi surface. A continuous scattering background up to high
frequencies is observed in all investigated high T$_c$ superconductors.
At temperatures below T$_c$ this scattering background becomes renormalized
for different frequencies below the energy gap $2\Delta(\vec{k})$, depending
on the scattering geometry.

\section{RESULTS AND DISCUSSION}
We investigated the temperature dependence of the electronic Raman
scat\-ter\-ing in single crystals of Bi$_2$Sr$_2$CaCu$_2$O$_{8+\delta}$ and
its variation with the oxygen content $\delta$. The crystals were
characterized by magnetic measurements in a SQUID magnetometer, by X-ray
diffraction, c-axis resistivity, and by Raman scattering. In order to change
the oxygen content $\delta$, we annealed the same single crystal subsequently
in Ar and in O$_2$. After each annealing step $\delta$ was determined by
comparing T$_c$ and the c-axis parameter to T$_c$($\delta$) and c($\delta$)
known from iodometric titration of polycrystalline samples. For the Raman
measurements we used the 488nm excitation line of an Ar$^+$-laser in
quasi-backscattering geometry and power levels below 15 W/cm$^2$.

\begin{figure}
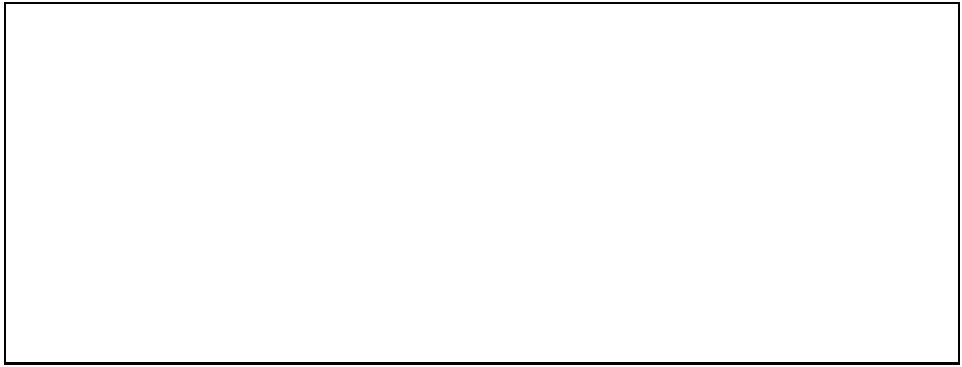

\framebox[5in]{\rule[.9in]{0in}{.9in}}
\caption{Normalized intensities of the A$_{1g}$, B$_{1g}$, and B$_{2g}$
symmetry component of the electronic Raman scattering for (a) $\delta$=0.17
and (b) $\delta$=0.29.}
\end{figure}

\begin{table}
\begin{tabular}{|c|c||c|c|c||c|}\hline
oxygen & T$_c$ & \multicolumn{3}{c||}{maximum (cm$^{-1}$) for} & maximum energy
gap \\
content $\delta$ & (K) & A$_{1g}$ & B$_{2g}$ & B$_{1g}$ &
$2\Delta_{max}/k_BT_c$ \\
\hline
0.17 & 86 & 330$\pm$20 & 370$\pm$30 & 520$\pm$20 & 8.7$\pm$0.3 \\ \hline
0.29 & 81 & 280$\pm$20 & 340$\pm$40 & 460$\pm$20 & 8.2$\pm$0.4 \\ \hline
\end {tabular}

table 1. Positions of the maximum for different symmetry components. The
maximum energy gap is determined by the B$_{1g}$ maximum.
\end{table}

In order to suppress phonon peaks, and to emphasize the redistribution of the
electronic Raman scattering intensity below T$_c$, spectra at T=10K are
divided by spectra at T=90K, see fig.\ 1. In the underdoped ($\delta$=0.17,
$\partial T_c/\partial\delta>0$) and in the overdoped regime ($\delta$=0.29,
$\partial T_c/\partial\delta<0$) the frequency behaviors for the different
symmetry components are consistent with a d$_{x^2-y^2}$-symmetry of the
energy gap according to ref.\ 2, i.e.\ the low frequency behaviors and the
maximum positions are different for the A$_{1g}$, B$_{2g}$, and B$_{1g}$
symmetry, see tab.\ 1. In the case of an energy gap with d$_{x^2-y^2}$-wave
symmetry, $2\Delta(\vec{k})$ has a different weight in a particular direction
of $\vec{k}$ for each symmetry component of the electronic Raman
scattering\cite{Devereaux}. This leads to the different frequency behaviors.
Regardless of the symmetry of the energy gap\cite{Devereaux}, the structure
in B$_{1g}$-symmetry can be identified with the maximum energy gap
$2\Delta_{max}$, see tab.\ 1. With doping the change of $2\Delta_{max}$ is
stronger than the change in T$_c$, indicating non-BCS behavior
($2\Delta_{max}/k_BT_c\neq const.$).

In fig.\ 2a we show the temperature dependence of the A$_{1g}$-  and
B$_{1g}$-peak of an overdoped Bi$_2$Sr$_2$CaCu$_2$O$_{8+\delta}$ single
crystal with $\delta$=0.27, T$_c$=83K. Both peaks show the same T dependence
up to T/T$_c$=0.73. This indicates that both peaks are due to the opening of
the superconducting energy gap. In fig.\ 2b the temperature dependence of the
B$_{1g}$-peak is shown for different oxygen contents $\delta$. Upon cooling
below T$_c$ the energy gap opens more rapidly in the underdoped crystal
($\delta$=0.17).

\begin{figure}
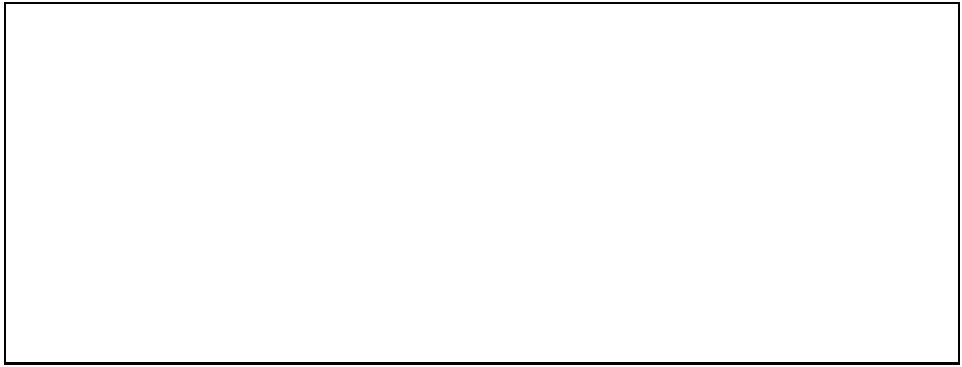

\framebox[5in]{\rule[.9in]{0in}{.9in}}
\caption{Temperature dependence of (a) the A$_{1g}$ and B$_{1g}$ peak for
$\delta$=0.27, T$_c$=83K and (b) the B$_{1g}$ peak for $\delta$=0.17 and
$\delta$=0.29 in Bi$_2$Sr$_2$CaCu$_2$O$_{8+\delta}$.}
\end{figure}

The results shown here are in good agreement with earlier measurements on
Bi$_2$Sr$_2$CaCu$_2$O$_{8+\delta}$ single crystals with different oxygen
concentrations\cite{Martin,Staufer}, explained by a change in anisotropy or
dimensionality, i.e coupling between CuO$_2$ planes with doping. However,
similar behavior is seen in the less anisotropic
YBa$_2$Cu$_3$O$_7$\cite{Hackl}. For this reason we emphasize the similarity
of this behavior with predictions based on paramagnon coupling. Within this
model spin fluctuations have a pair-breaking and pair-binding effect. The
opening of the superconducting energy gap leads to a decrease of low
frequency spin fluctuations and thus to less pair-breaking. This feedback
effect leads in underdoped samples to a more abrupt temperature dependence of
$\Delta(T)/\Delta_{max}$ and to a higher magnitude of $2\Delta_{max}$
compared to BCS theory\cite{Pao} and is in good agreement with our results.
Since with increased oxygen content $\delta$ the
Bi$_2$Sr$_2$CaCu$_2$O$_{8+\delta}$ single crystals are farther away from the
antiferromagnetic insulator, this feedback effect should become reduced. This
explains the smaller $2\Delta_{max}/k_BT_c$ and its weaker temperature
dependence for the overdoped crystals compared with the underdoped ones.

In conclusion, the frequency dependencies of the different symmetry
components of the electronic Raman scattering are consistent with a
d$_{x^2-y^2}$-wave energy gap. Furthermore the measured magnitude of the
energy gap, its temperature dependence and its variation with the oxygen
content $\delta$ is consistent with a pairing mechanism due to
antiferromagnetic spin fluctuations, i.e. paramagnon coupling.

We would like to thank D.\ Einzel, T.P.\ Devereaux, A.\ Kampf, M.\ Krantz,
M.\nolinebreak[4]\ Cardona, and K.\ Maki for stimulating discussions. This
work was supported by DFG through SFB 341.


\begin{thebibliography}{9}

\bibitem{Schrieffer} J.R.\ Schrieffer, {\it Solid State Commun.}\ {\bf 92},
129 (1994) and references therein.

\bibitem{Devereaux} T.P.\ Devereaux, D.\ Einzel, et al.,
{\it Phys.\ Rev.\ Lett.}\ {\bf 72}, 369 (1994); T.P.\ Devereaux and
D.\ Einzel, to be published (Sissa preprint \# cond-mat/9408019).

\bibitem{Pao} C.H.\ Pao and N.E.\ Bickers,
{\it Phys.\ Rev.\ Lett.}\ {\bf 72}, 1870 (1994); P.\ Monthoux and
D.\ Scalapino, {\it Phys.\ Rev.\ Lett.}\ {\bf 72}, 1874 (1994).

\bibitem{Martin} M.\ Boekholt et al., {\it Physica C} {\bf 175}, 127 (1991).

\bibitem{Staufer} T.\ Staufer et al., {\it Phys.\ Rev.\ Lett.}\ {\bf 68},
1069 (1992).

\bibitem{Hackl} R.\ Hackl et al., {\it Phys.\ Rev.\ B}\ {\bf 38}, 7133
(1988).

\end{thebibliography}
\end{document}